\documentclass[prb,superscriptaddress,twocolumn,showpacs,preprintnumbers,amsmath,floatfix]{revtex4}
\usepackage{amssymb}
\usepackage{graphicx}

\begin{document}

\title{Interaction-induced metallic state in graphene on hexagonal boron nitride}
\author{Jin-Rong Xu}
\affiliation{Shanghai Key Laboratory of Special Artificial Microstructure Materials and Technology, \\
School of Physics Science and Engineering, Tongji University, Shanghai 200092, P.R. China}
\affiliation{School of Mathematics and Physics, Anhui Jianzhu University, Hefei, Anhui, 230601, P.R. China}
\author{Ze-Yi Song}
\affiliation{Shanghai Key Laboratory of Special Artificial Microstructure Materials and Technology, \\
School of Physics Science and Engineering, Tongji University,
Shanghai 200092, P.R. China}
\author{Chen-Guang Yuan}
\affiliation{Shanghai Key Laboratory of Special Artificial Microstructure Materials and Technology, \\
School of Physics Science and Engineering, Tongji University,
Shanghai 200092, P.R. China}
\author{Yu-Zhong Zhang}
\email[Corresponding author. ]{Email: yzzhang@tongji.edu.cn}
\affiliation{Shanghai Key Laboratory of Special Artificial Microstructure Materials and Technology, \\
School of Physics Science and Engineering, Tongji University, Shanghai 200092, P.R. China}
\affiliation{Beijing Computational Science Research Center, Beijing 100084, P.R. China}
\date{\today }

\begin{abstract}
The Coulomb interaction is widely known to enhance the effective mass of interacting particles and therefore tends to favor a localized state at commensurate filling. Here, we will show that, in contrast to this consensus, in a van der Waals heterostructure consisting of graphene and hexagon boron nitride (h-BN), the onsite Coulomb repulsion will at first destroy the localized state. This is due to the fact that the onsite Coulomb repulsion tends to suppress the asymmetry between neighboring carbons induced by h-BN substrate. We corroborate this surprising phenomenon by solving a tight-binding model with onsite Coulomb repulsion treated within coherent potential approximation, where hopping parameters are derived from density functional theory calculations based on the graphene/h-BN heterostructure. Our results indicate that both gapless and gapped states observed experimentally in graphene/h-BN heterostructures can be understood after a realistic value of the onsite Coulomb repulsion as well as different interlayer distances are taken into account. Finally, we propose ways to enhance the gapped state which is essential for potential application of graphene to next-generation electronics. Furthermore, we argue that band gap suppressed by many-body effect should happen in other van der Waals heterostructures.
\end{abstract}


\maketitle

\section{Introduction~\label{intro}}

Graphene, a Dirac semimetal, is considered as a potential candidate for replacing silicon in next-generation electronics provided large gaps of
order 300~K can be opened at charge neutral point. Various ways have been used to realize the gap
opening~\cite{Castro2007,Han2007,Elias2009,Ci2010,Hunt2013}. Among those, heterostructure consisting of two-dimensional layers of graphene and
atomically flat hexagonal boron nitride (h-BN) coupled by van der Waals interactions attracts considerable
interests~\cite{Hunt2013,Xue2011,Britnell2012,Slawinska2010,Dean2011,Zhong2011,Gao2012,Amet2013,Tang2013,San-Jose075428,Titov2014,Wijk2014,Abergel2015,Slotman2015}.

While it was predicted by density functional theory (DFT) calculations that a band gap of $53$ meV can be generated in the van der Waals
heterostructure~\cite{Giovannetti2007}, such an insulating state was not always detected in experiments~\cite{Hunt2013, Xue2011,Amet2013,Usachov2010,Roth2013,Chen2014,Sediri2015}.
The discrepancy is originally ascribed to the lattice mismatch of 1.8\%  between graphene and h-BN which leads to moir\'{e} patterns in real space with periodicity determined by interlayer twist angle~\cite{Xue2011,Roth2013,Titov2014,Sachs2011}. This is due to the fact that nonzero twist may give rise to a restoration of sublattice inversion symmetry on spatial average, possibly leading to a metallic state~\cite{Xue2011}. However, such a metallic state is not anticipated by the noninteracting model, where a small gap persists even if lattice mismatch is involved~\cite{Kindermann2012}. Also the insulating state with large band gap of around 100-300K observed experimentally~\cite{Hunt2013} can not be accounted for by the noninteracting model~\cite{Kindermann2012}. Then, a natural speculation arises that electronic interactions may enhance the gapped state. Indeed, various experiments indicate the importance of many-body effect in the heterostructure~\cite{Dean2011,Chen2014,Dean2013,Shi2014}.

Recently, long-range Coulomb interactions are studied at either weak coupling limit or strong coupling limit~\cite{Song2013,Bokdam2014,Jung2015} which confirm the speculation. But it is known from constrained random phase approximation calculations that the dominant onsite Coulomb repulsion $U^*$ is reduced by a weighted average of nonlocal interactions $\bar{V}$, resulting in an effective on-site Coulomb interaction of $U=U^*-V=1.6t$, where $t$ is the hopping integral of $\pi$ electrons between nearest neighbor carbons~\cite{Schueler2013}. Therefore, it's better to first understand the role of the effective on-site Coulomb interaction which is of intermediate strength with respect to the kinetic energy of the $\pi$ electrons.

In this paper, we will show that the onsite Coulomb repulsion will first suppress the band gap in the graphene/h-BN heterostructure even in the absence of lattice mismatch. The underlying physics behind the unexpected phenomenon comes from competition between two types of insulating states, namely the band insulator and the Mott insulator. As is well known, at $U=0$, lattice inversion symmetry of graphene is broken by coupling to the h-BN substrate,
leading to a charge imbalance between two neighboring carbons and a band gap at half-filling (i.e. the charge neutral point).
However, as $U$ is switched on, the onsite Coulomb repulsion tends to localize the electrons on graphene and thus
suppress the hybridization between graphene and h-BN. Therefore, the asymmetry between neighboring carbons as well as the corresponding band gap,
induced by interlayer coupling, will be suppressed. Further increasing the onsite Coulomb repulsion will drive the system into the Mott insulating state.

Besides the reduction of the band gap induced by the onsite Coulomb repulsion of intermediate strength, variation of interlayer distance, which dominates the interlayer coupling between graphene and h-BN, will also strongly affect the band gap. Since large variance of the interlayer distance may exist among different measurements due to the weak van der Waals interaction and diverse experimental conditions, and the onsite Coulomb repulsion inevitably exists in real materials, our findings may provide complementary ways to understand the conundrum of why both gapped and gapless states can be observed in different experiments. We corroborate the above findings by solving a tight-binding model with onsite Coulomb repulsion treated within coherent potential approximation. The hopping parameters of the tight-binding model are derived from DFT calculations based on the graphene/h-BN heterostructure.

The paper is organized as follows. In Sec.~\ref{MM}, we describe the details of the structure, the model, and the method we used. In Sec.~\ref{res}, we present our results including density of states (DOS), gap, difference in occupation numbers between neighboring carbons, total self-energy, spectral function, and phase diagram in the plane of onsite Coulomb interaction and interlayer distance. Corresponding indications of our results to experiments are also discussed in this section. We summarize our findings in Sec.~\ref{conclusion}.

\section{Model and Methods~\label{MM}}

\subsection{Details of the heterostructures~\label{cartoon}}

The graphene/h-BN heterostructure we studied are composed of one h-BN monolayer and one graphene monolayer coupled by van der Waals interaction. Since the physical picture of interaction-induced reduction of the band gap is independent of lattice structure, for simplicity, we only study the van der Waals heterostructure without lattice mismatch. Further advantage to use lattice-matched structure is that, as long as the gapless state appears, it can be solely attributed to the inclusion of on-site Coulomb repulsion, since possible reduction of the band gap due to lattice mismatch is completely precluded.

There are three stable and metastable configurations of the heterostructure as shown in the insets of Fig.~\ref{Fig:two} and Fig.~\ref{Fig:five}, respectively. The most stable configuration with lowest total energy is called AB-N stack where one carbon is above boron and the other centered above a h-BN hexagon as seen in the insets of Fig.~\ref{Fig:two}. Two metastable configurations are given in the insets of Fig.~\ref{Fig:five}, denoted by AB-B stack with one carbon over nitrogen and the other centered above a h-BN hexagon and AA stack with one carbon over boron and the other over nitrogen, respectively. Unless specified, the lattice structure we used in our paper is of AB-N stack.

\subsection{Tight-binding model~\label{TBmodel}}

The tight-binding model with onsite Coulomb interaction used to describe the graphene/h-BN heterostructure is given by
\begin{equation}
\label{fullmodel}
H=\sum_{i,j\in C,B,N,\sigma }t_{ij}a_{i\sigma }^{\dag }a_{j\sigma
}+U\sum_{i\in C}(n_{i\uparrow }-1/2)(n_{i\downarrow }-1/2)
\end{equation}%
where $a_{j\sigma }$ ($a_{i\sigma }^{\dag }$) is the annihilation (creation) operator of an electron with spin $\sigma $ at site j (i),
and $n_{i\sigma}=a_{i\sigma }^{\dag }a_{i\sigma }$ the density operator. $i,j\in C,B,N$ denotes the summation over the sites on both
graphene and h-BN while $i\in C$ the graphene only. $t_{ij}$ is the hopping integral between site i and j derived from DFT calculations and $U$ is the effective onsite Coulomb repulsion. We are interested in the half-filled case, which
is corresponding to the charge neutral point in experiments. Since no long-range magnetic order is observed in the heterostructure experimentally, we focus on the paramagnetic state throughout the paper.

\subsection{Details of band structure calculations~\label{DFT}}

In order to determine the hopping integrals $t_{ij}$ in model~(\ref{fullmodel}), we first perform DFT calculations within local density approximation (LDA) to obtain band structure of the graphene/h-BN heterostructure, based on projector augmented wave (PAW) method~\cite{Bloechl1994}, as implemented in VASP code~\cite{Kresse1993,Kresse1996}. The convergence of total energy with respect to K-point sampling, cutoff energies and vacuum distance between neighboring heterostructures has been carefully examined. The plane wave cutoff energy of 800~eV, a $\Gamma$-centered K-point grid of $33\times33\times1$ and a vacuum distance of 17 \AA ~are chosen when lattice constants and atomic positions of lattice-matched graphene/h-BN heterostructures are fully relaxed, with residual forces on each atom less than 1~meV/\AA. Since the distances between graphene and h-BN may vary considerably among different experiments due to the weak van der Waals coupling and diverse experimental conditions, the interlayer distances will be treated as a tuning parameter while other structural parameters are fixed according to the lattice optimization. We use $41\times41\times1$ Monkhorst-Pack k-point meshes~\cite{Monkhorst} to integrate over the Brillouin zone in order to precisely determine the Fermi energy. With the parameters described above, we can reproduce the band structures of Ref.~[\onlinecite{Zhong2011}] and Ref.~[\onlinecite{Giovannetti2007}] where different equilibrium interlayer distances are used.

\begin{figure}[htbp]
\includegraphics[width=0.48\textwidth]{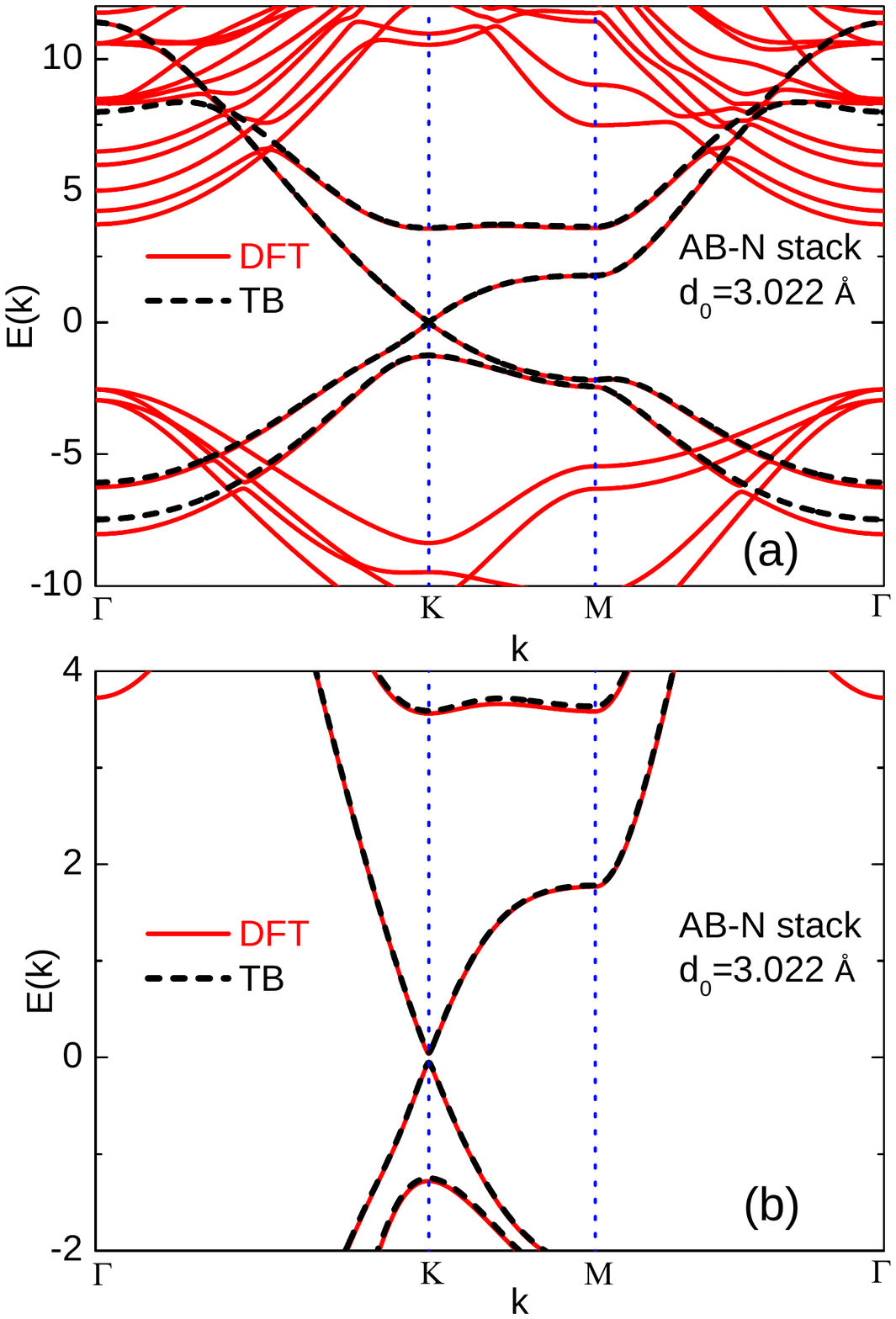}
\caption{(Color online) Comparison between band structures of graphene/h-BN heterostructure calculated from density functional theory (DFT) and the derived tight-binding model~(\ref{fullmodel}) at $U=0$ for AB-N stack with interlayer distance of $d_0=3.022$~\AA. (b) is a blowup of (a) around the Fermi level. Solid (red) lines are the band structure from DFT calculations while dashed (black) ones from the derived tight-binding model.}
\label{Fig:supplethree}
\end{figure}

Then, the hopping integrals can be derived through transformation from Bloch space to maximally localized Wannier functions (MLWFs) basis by using wannier90 code~\cite{Marzari2012,Mostofi}. Four bands close to the Fermi energy which are mainly contributed from $p_z$ orbitals of two carbons of different sublattices, one nitrogen, and one boron are taken into account. Comparisons between the band structures calculated from DFT and the derived tight-binding model~(\ref{fullmodel}) at $U=0$ are shown in Fig.~\ref{Fig:supplethree} for AB-N stack with interlayer distance of $d_0=3.022$~\AA. Only a little differences around the $\Gamma$ point due to strong band entanglement can be seen, which will not affect the applicability of our results to the heterostructures as those states are far away from the Fermi level and therefore irrelevant to the band gap. Please note, as the noninteracting part in Eq. (\ref{fullmodel}) are derived from DFT calculations, the effect of nonlocal terms of the Coulomb interactions has been implicitly incorporated in our study at the DFT level. In this paper, we will focus on the effect of the optimal onsite Coulomb interaction beyond the DFT level~\cite{CommentTB}.

\subsection{Coherent potential approximation to the model~(\ref{fullmodel})~\label{CPA}}

In order to verify the above scenario of interaction-induced metallization and its relevance to graphene/h-BN heterostructure, we employ coherent potential
approximation (CPA)~\cite{Elliot1974,Jarrell2001} to solve the tight-binding model~(\ref{fullmodel}). By applying the alloy analogy approach~\cite{Hubbard1963}, the interacting system can be viewed as a disordered alloy where an electron with spin $\sigma$ moving on the graphene layer encounters either a potential of $U/2$ at a site with a spin $-\sigma$ present or $-U/2$ without. Then, the model Hamiltonian (\ref{fullmodel}) is replaced by a one-particle Hamiltonian with disorder potential which is of the form

\begin{equation}
H=\sum_{i,j\in C,B,N,\sigma }t_{ij}a_{i\sigma }^{\dag }a_{j\sigma}+\sum_{i\in C,\sigma} E_{i\sigma} n_{i\sigma},
\label{Randommdoel}
\end{equation}
where the disorder potential $E_{i\sigma}$ satisfies
\begin{equation}
E_{i\sigma}=\left\{
\begin{array}{lccl}
U/2& &\text{with probability}&  \langle n_{i\bar{\sigma}} \rangle\\
-U/2& &\text{with probability}& 1-\langle n_{i\bar{\sigma}}\rangle
\end{array}\right. .
\label{rand.potential}
\end{equation}
Here, $\langle n_{i\sigma} \rangle$ is the average occupancy of electrons on site $i$ of the graphene layer with spin $\sigma$. The Green's function corresponding to the one-particle Hamiltonian has to be averaged over all possible disorder configurations. The averaging can not be performed exactly. To solve the alloy problem, the coherent potential approximation (CPA) is used~\cite{Elliot1974,Jarrell2001}, where the disorder potential $E_{i\sigma}$ is replaced by a local complex and energy-dependent self-energy. The details of the CPA method applied to the model (\ref{fullmodel}) are given in the appendix~\ref{GF}. Here, we should stress that, although above treatment itself has a few shortcomings~\cite{Gebhard1997}, it remains valuable as a computationally simple theory capable of capturing the Mott metal-insulator transition of many-body systems. For example, it successfully reproduces the phase diagram of ionic Hubbard model at half filling~\cite{Hoang2010}. Moreover, the critical value of the Mott transition of $U_c/t \approx 3.5$~\cite{RowlandsCPB2014} obtained within CPA based on a Hubbard model on honeycomb lattice at half filling is in excellent agreement with the quantum Monte Carlo simulations~\cite{Assaad2013,Sorella2012,Toldin2015} and cluster dynamical mean field theory calculations~\cite{Wu2010,Liebsch2011,CommentAF}. More discussions about CPA can be found in Refs.~[\onlinecite{RowlandsJPCM1,RowlandsJPCM2,CommentCPA}].

\section{results~\label{res}}

\begin{figure}[htbp]
\includegraphics[width=0.48\textwidth]{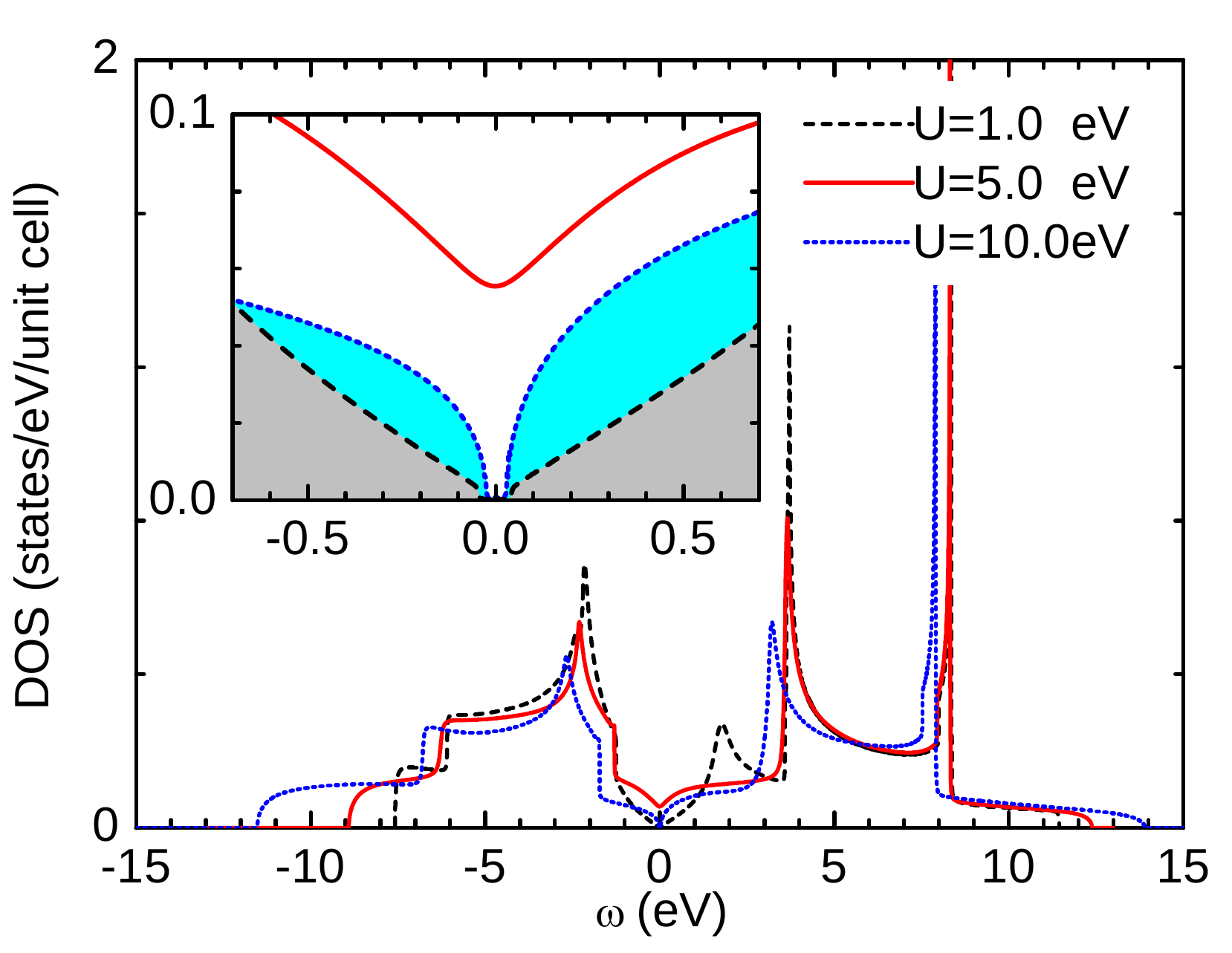}
\caption{(Color online) Densities of states (DOS) for three different values of $U$ with interlayer distance of $d_0=3.022$~\AA. Inset is the blow-up of DOS around the Fermi Level.}
\label{Fig:one}
\end{figure}

Fig.~\ref{Fig:one} shows DOSs at three different values of $U$. Here, an equilibrium interlayer distance of $d_0=3.022$ \AA~
between graphene and h-BN is used according to Ref.~\onlinecite{Zhong2011}. It is found that at a small value of $U=1.0$ eV, DOS at the Fermi level remains
zero which is similar to the noninteracting case. However, when $U$ becomes larger, for example at $U=5.0$ eV, finite DOS appears at the charge neutral
point, indicating emergence of an unconventional metallic state induced by onsite electronic Coulomb interaction. Further increasing $U$ will again result
in a gapped state as evident from the DOS at $U=10.0$ eV.

\begin{figure}[htbp]
\includegraphics[width=0.46\textwidth]{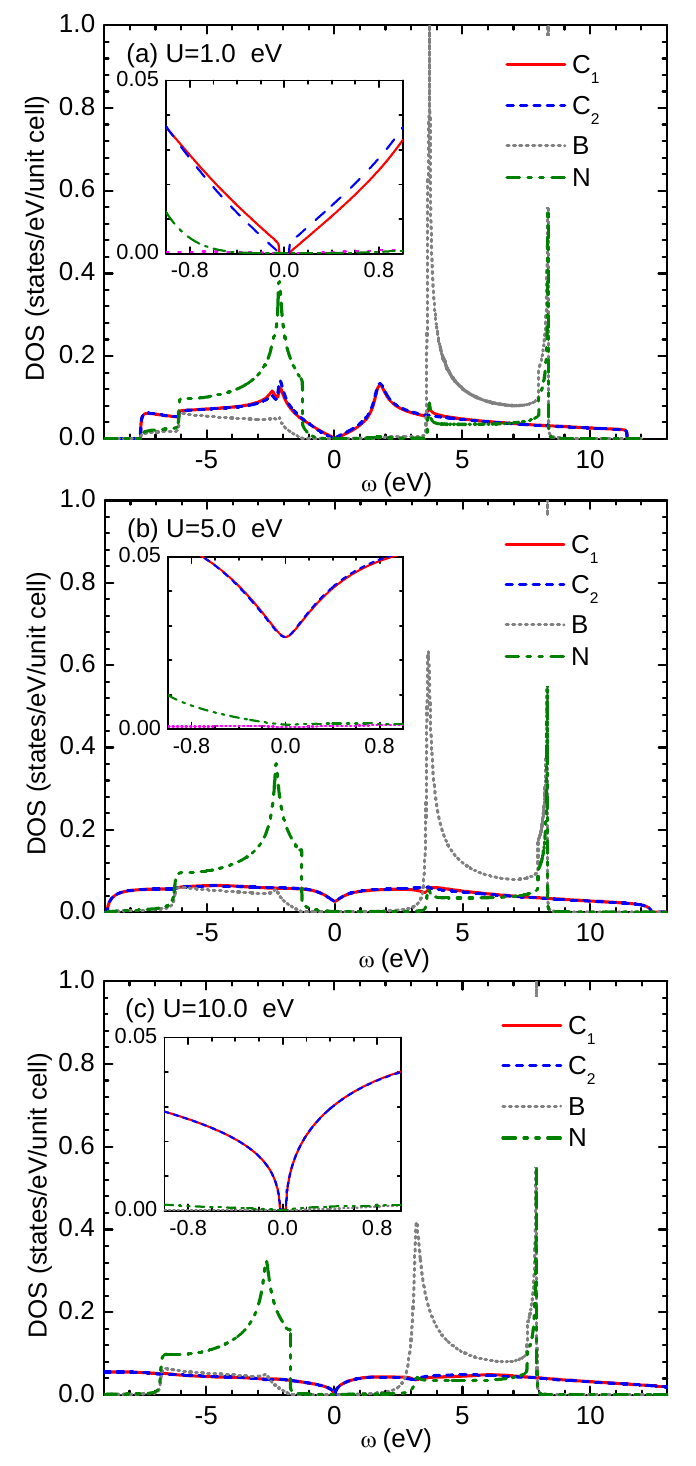}
\caption{(Color online) Partial Densities of states (DOS) for three different values of $U$ with interlayer distance of $d_0=3.022$~\AA.
(a) $U=1.0$~eV, (b) $U=5.0$~eV, and (c) $U=10.0$~eV. Insets are the blow-ups of DOS around the Fermi Level. C$_1$ and C$_2$ are the carbons of different sublattices. B and N denote boron and nitrogen, respectively.}
\label{Fig:new}
\end{figure}

The partial DOS is presented in Fig.~\ref{Fig:new} for above three different values of $U$. It is found that in all the cases, DOS around the Fermi level are mainly contributed from both carbons C$_1$ and C$_2$. The contributions of boron and nitrogen to the DOS around the Fermi level are negligibly small, compared to those of carbons. Moreover, while small difference between the DOS of C$_1$ and C$_2$ can be found at $U=1.0$~eV, the difference becomes indiscernible at $U=5.0$~eV and $U=10.0$~eV. This implies that the asymmetry between carbons induced by the substrate h-BN is suppressed by the onsite Coulomb repulsion. Due to the effective restoration of the inversion symmetry between carbons, the band gap induced by the asymmetry gradually vanishes, and finally an interaction-induced metallic state appears.

\begin{figure}[htbp]
\includegraphics[width=0.48\textwidth]{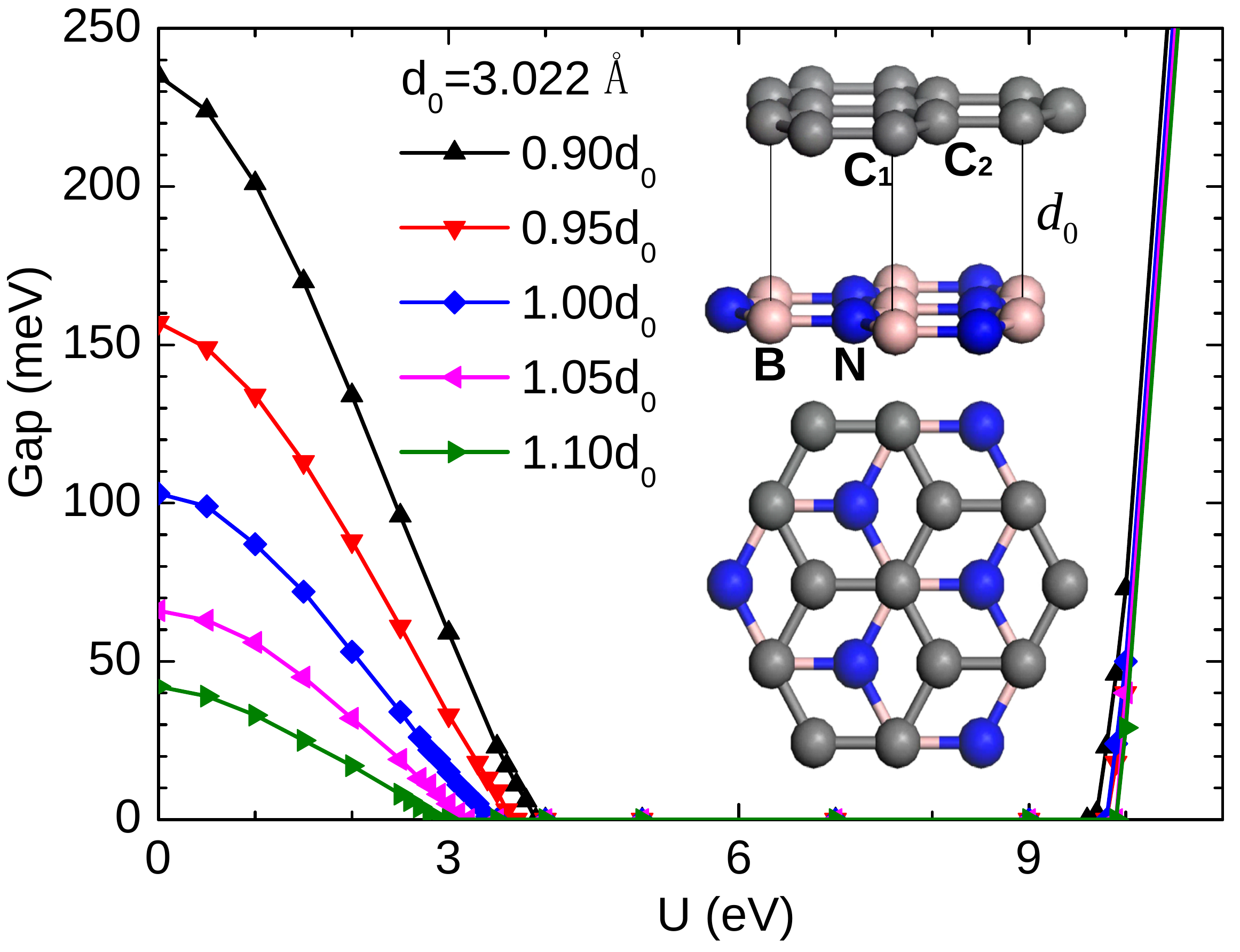}
\caption{(Color online) Evolution of gap size as a function of onsite Coulomb repulsion for several different interlayer distances in the unit of equilibrium distance $d_0=3.022$~\AA. Insets show the most stable lattice configuration with lowest total energy, called AB-N stack where one carbon is above boron and the other centered above a h-BN hexagon. upper panel is side view while lower one is top view. C$_1$ and C$_2$ are the carbons of different sublattices. B and N denote boron and nitrogen, respectively.}
\label{Fig:two}
\end{figure}

The evolution of gap size as a function of onsite Coulomb repulsion is shown in Fig.~\ref{Fig:two} for several different interlayer distances in the unit of equilibrium distance $d_0=3.022$~\AA.
While at weak interacting region, both the gap amplitudes and the critical values where the gap vanishes are strongly dependent on the interlayer
distances, implying that insulating behavior in this region is dominated by coupling between graphene and h-BN, at large interacting region,
those quantities remain almost unchanged, indicating that the gapped state at large $U$ is probably irrelevant to the h-BN substrate. Moreover,
the strong variance of critical values at small $U$ region as a function of the interlayer distances also suggests that such an interaction-driven
insulator-to-metal transition can be tuned by varying the interlayer distances between graphene and h-BN, which can be easily realized by applying
out-of-plane strain to the van der Waals heterostructure.

\begin{figure}[htbp]
\includegraphics[width=0.48\textwidth]{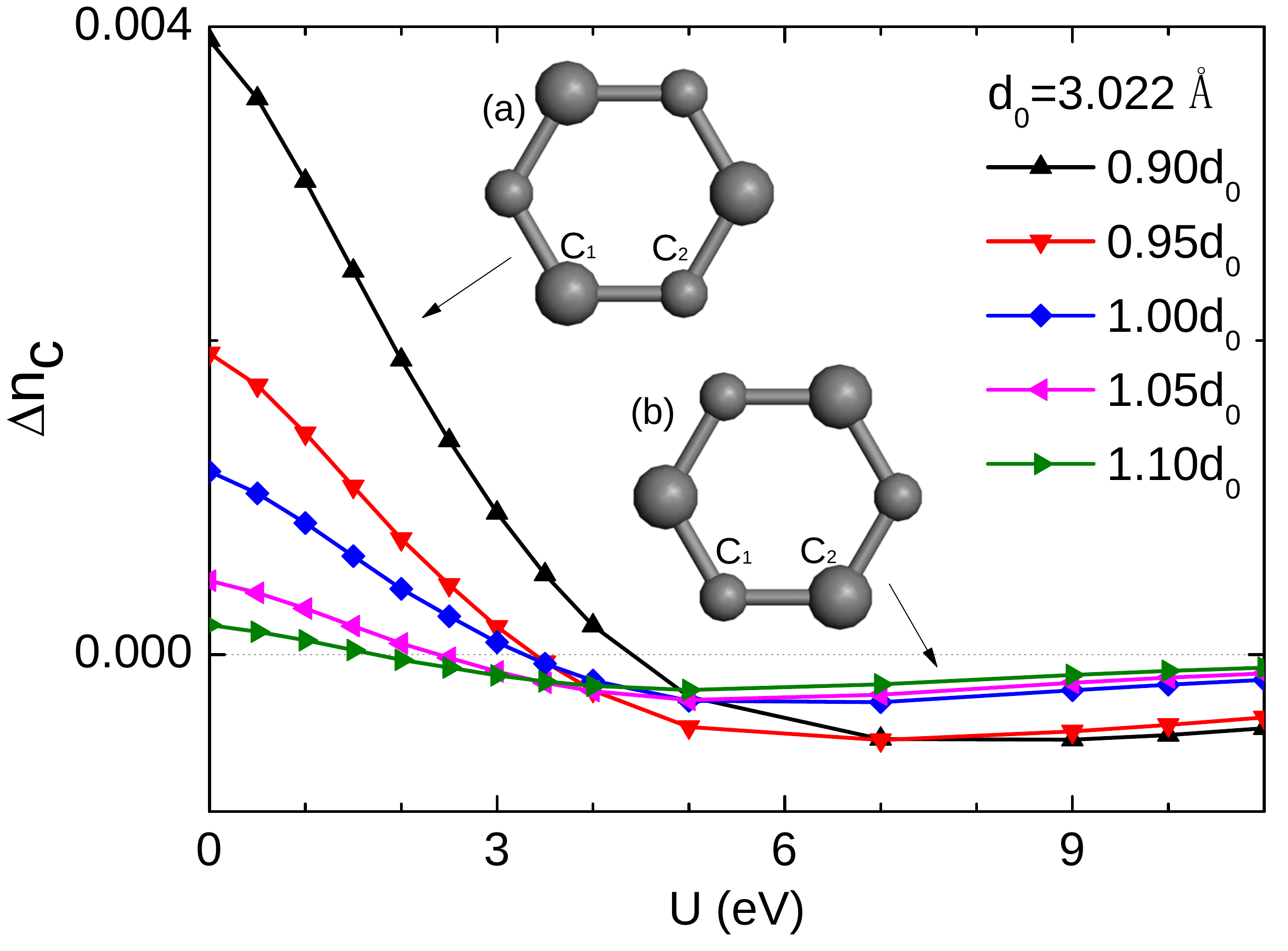}
\caption{(Color online) Difference of occupation number between two nearest neighbor carbons $\Delta n_c$ as a function of $U$ for several different interlayer distances in the unit of equilibrium distance $d_0=3.022$~\AA. Insets are cartoons for the graphene layer where larger dots denote charge-richer sites while smaller ones indicate charge-poorer sites. (a) and (b) represents for the case of smaller U region where interlayer hybridizaions play dominant roles and larger U region where local potentials dominate the charge distributions, respectively. C$_1$ and C$_2$ are the carbons of different sublattices.}
\label{Fig:three}
\end{figure}

\begin{figure}[htbp]
\includegraphics[width=0.48\textwidth]{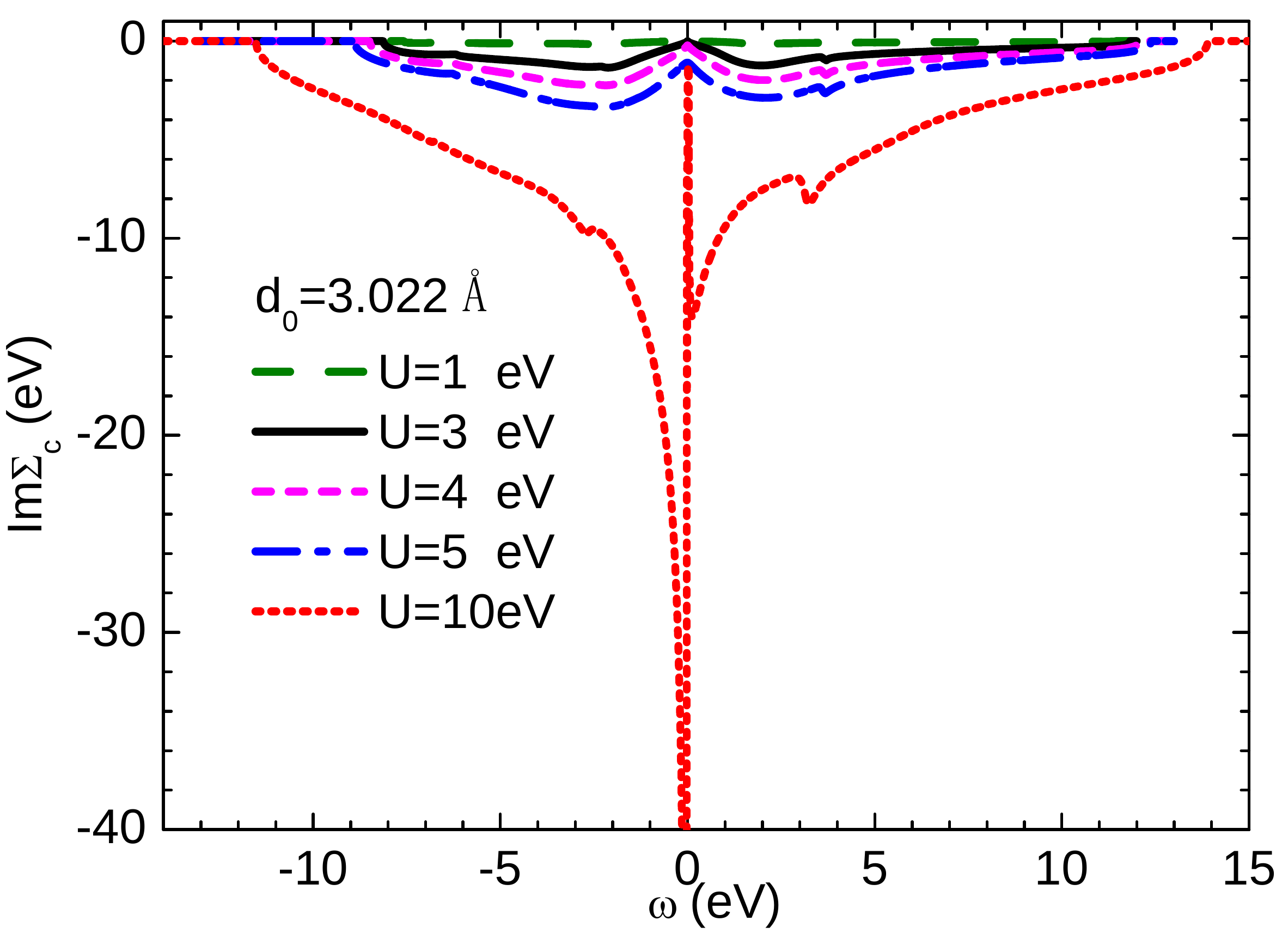}
\caption{(Color online) Imaginary part of total self-energy of two neighboring carbons $Im \Sigma_C$ for several values of $U$ with interlayer distance of $d_0=3.022$~\AA.}
\label{Fig:four}
\end{figure}

In order to reveal the nature of these two insulating states, difference of occupation number between two nearest neighbor carbons $\Delta n_c$
and imaginary part of total self-energy of two neighboring carbons $Im \Sigma_C$ as a function of $U$ are investigated as shown in
Figs.~\ref{Fig:three} and~\ref{Fig:four}, respectively. It is found that $\Delta n_c$ is rapidly suppressed as $U$ increases, indicating that
the symmetry breaking between two neighboring carbons induced by h-BN substrate is effectively restored by the interaction continuously, leading
to a reduction of gap amplitude as detected in Fig.~\ref{Fig:two}. Combining the fact that the imaginary part of total self-energy at small
$U$, as displayed in Fig.~\ref{Fig:four} for $U=1$ or $3$~eV, vanishes in the vicinity of the charge neutral point, we conclude that
in the small interacting region, it is still a band insulator where symmetry breaking dominates the gap opening. However, the situation is
completely changed at large $U$ where the imaginary part of total self-energy is divergent within the gap, as seen in Fig.~\ref{Fig:four}
at $U=10$~eV, which is a characteristic feature of a Mott insulator~\cite{Georges1996}. In the intermediate range of onsite Coulomb interaction, the imaginary part of total self-energy, which will cause band broadening, becomes larger, (see $U=4$ or $5$ in Fig.~\ref{Fig:four}), while the asymmetry
between neighboring carbons becomes smaller as indicated by smaller charge difference in Fig.~\ref{Fig:three}, compared to the weak interacting case.
A metallic state induced by many-body interaction appears, since the small asymmetry can not provide large enough separation between broadened conduction and valence bands.

Another interesting phenomenon presented in Fig.~\ref{Fig:three} is that the difference of occupation number between neighboring carbons
$\Delta n_c$ changes from positive to negative value, indicating a reversal from charge-richer to charge-poorer site in graphene layer
for all different interlayer distances. This can be ascribed to two competing effects induced by h-BN substrate. One is the substrate-induced
different local potentials between neighboring carbons, and the other is the difference of interlayer hybridizations between different
carbons and the substrate. We find that the carbon above boron has higher local potential and stronger interlayer hybridization than
that sitting above the center of h-BN hexagon. When $U$ is small, more electrons prefer to occupy the carbon above boron since it
can gain more kinetic energy through the interlayer hopping. However, when $U$ becomes larger, the electrons in graphene get more
localized, and the interlayer hoppings are suppressed. As a result, more electrons choose to occupy the carbon above the center of h-BN hexagon
due to the lower local potential. Such a physical picture can also be applied to understand why a reversal from charge-richer to
charge-poorer site happens at fixed value of $U$ as the interlayer distance is varied (See e.g. $U=3$ or $4$~eV in Fig.~\ref{Fig:three}).
Due to the competition between site-dependent local potentials and interlayer hybridizations, smaller interlayer distance results in larger
interlayer hybridizations, and thus favors more electrons on the carbon above boron in order to gain more kinetic energy, while larger
interlayer distance corresponds to weaker interlayer hybridizations and therefore less electrons on the carbon above boron due to the
higher local potential.

\begin{figure}[htbp]
\includegraphics[width=0.48\textwidth]{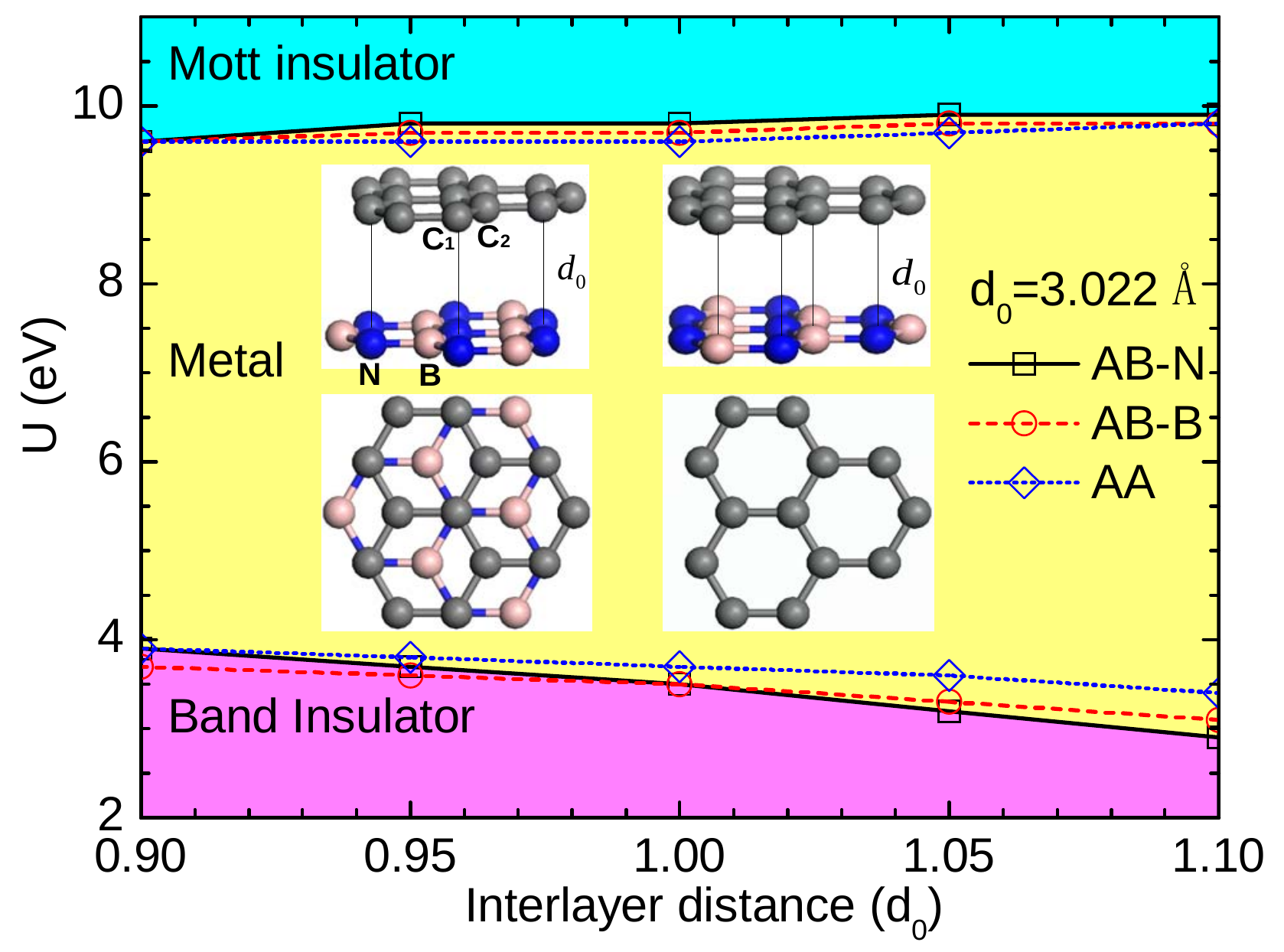}
\caption{(Color online) Phase diagram in the plane of interaction in the unit of eV and interlayer distance in the unit of equilibrium distance $d_0=3.022$~\AA. Lines are guidance for eyes. Insets are two metastable lattice configurations denoted by AB-B stack with one carbon over nitrogen and the other centered above a h-BN hexagon (left two panels), and AA stack with one carbon over boron and the other over nitrogen (right two panels), respectively. Upper panels are side views while lower panels are top views. The most stable lattice configuration is shown in the insets of Fig.~\ref{Fig:two}. C$_1$ and C$_2$ are the carbons of different sublattices. B and N denote boron and nitrogen, respectively.}
\label{Fig:five}
\end{figure}

The resulting phase diagram in the plane of interaction and interlayer distance is shown in Fig.~\ref{Fig:five}. Here, we present not only the
phase boundaries for most stable lattice configuration, but also those for two metastable configurations with one carbon above nitrogen and the
other above the center of h-BN hexagon represented by AB-B stack, and with one carbon above boron and the other above nitrogen denoted by AA stack, respectively. We find that consecutive phase transitions from band insulator to metal and then to Mott insulator always happen despite differences in lattice configurations and the interlayer distances, indicating that the interaction-driven metallization should always occur even if lattice mismatch is taken into account.

Finally, let us discuss the implication of our results to the experimental conundrum where metallic and insulating states are both observed in different measurements~\cite{Xue2011,Hunt2013,Amet2013,Usachov2010,Roth2013,Chen2014,Sediri2015}. Although it was initially argued that nonzero twist may give rise to a restoration of sublattice inversion symmetry on spatial average, leading to a metallic state~\cite{Xue2011}, it was found from transport measurements that the gap remains large, e.g. of the order 200~K, even if the twist angle $\delta$ becomes larger than $2^{\circ}$~\cite{Hunt2013}. On the contrary, no gap was observed in another experiment when twist angle $\delta$ is larger than $1^{\circ}$~\cite{Woods2014}. Similar inconsistency also happens between the transport measurements and angular resolved photoemission spectroscopic studies (ARPES). From ARPES, a band with linear dispersion is observed crossing the Fermi level in the heterostructure with moir\'{e} wavelength of 9~nm, clearly indicating a metallic state~\cite{Roth2013}, which is in contrast to the gapped state with same moir\'{e} wavelength reported by transport measurements~\cite{Hunt2013}. These discrepancies indicate that existing explanations of the gapless or gapped state observed in different experiments based on lattice relaxations and misorientations~\cite{Jung2015,Woods2014} can not resolve the conundrum completely.

\begin{figure}[htbp]
\includegraphics[width=0.48\textwidth]{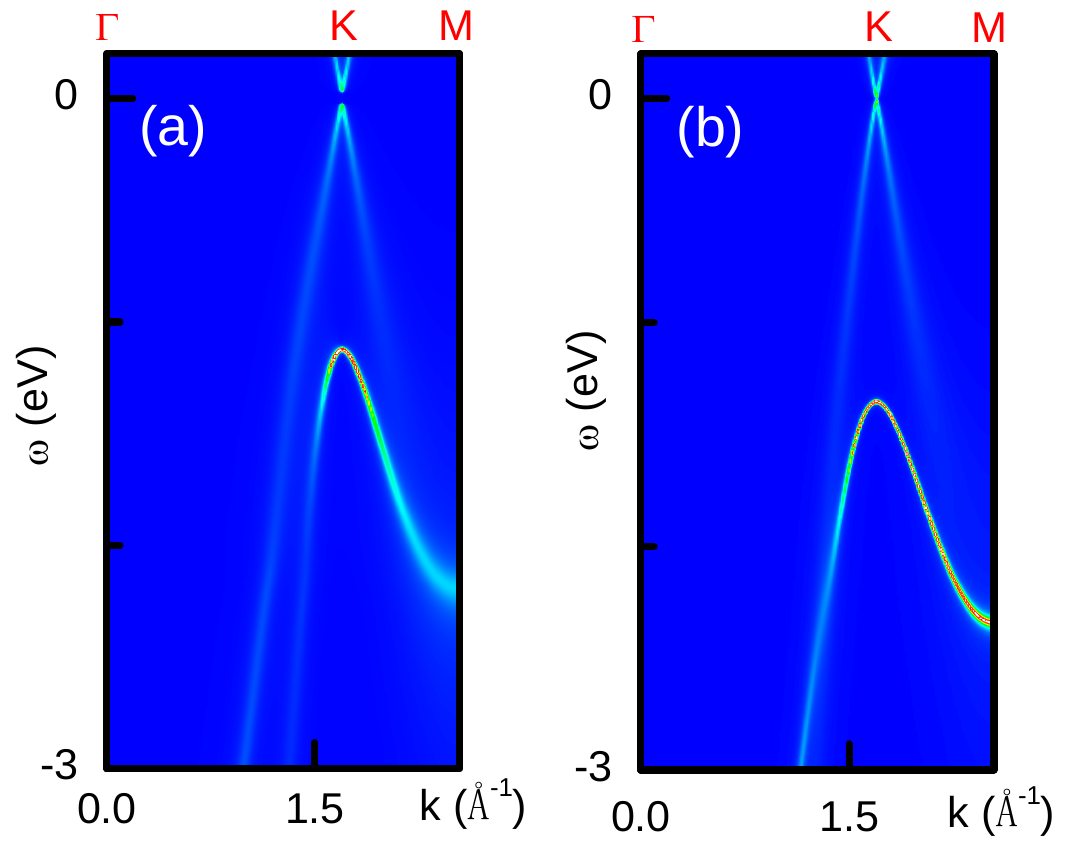}
\caption{(Color online) (a) and (b) are the spectral functions along $\Gamma-K-M$ line in the Brillouin zone at $U=3$~eV for two different interlayer distances of $0.9d_0$  and $1.1d_0$ with $d_0=3.022$~\AA, respectively. }
\label{Fig:six}
\end{figure}

Here, according to the phase diagram, we propose a complementary way to fully understand the contradictory experiments based on the existence of onsite Coloumb interaction and the difference of interlayer distance, both of which are unavoidable in real materials, in addition to the way of considering commensurate-incommensurate phase transition~\cite{Woods2014}. After the onsite Coulomb repulsion is taken into account, if we adopt $U=3$~eV, which is of the same value used for bilayer graphene~\cite{Zhang2016}, we can find from Fig.~\ref{Fig:five} that the heterostructure is just located in the vicinity of the phase boundary between band insulator and metal. Therefore the electronic property of the heterostructure is susceptible to the variation of interlayer distance which may be considerably large among different experiments due to the weak van der Waals interaction between graphene and h-BN. In Fig.~\ref{Fig:six}, we further show the spectral functions along $\Gamma-K-M$ line in the Brillouin zone at $U=3$~eV for two different interlayer distances of $0.9d_0$ (see Fig.~\ref{Fig:six} (a)) and $1.1d_0$ (see Fig.~\ref{Fig:six} (b)) with $d_0=3.022$~\AA. While a small gap of $59$~meV is clearly visible at the distance of $0.9d_0$, nearly linear dispersive band crossing the Fermi level appears at the distance of $1.1d_0$ as observed in the ARPES studies~\cite{Roth2013}.

According to the phase diagram, we further propose that applying stress perpendicular to the graphene plane may open a large band gap as required for making logic transistors. Another way to enhance the gap is to put one more h-BN on top of graphene which will enhance the asymmetry between neighboring carbons and therefore lead to a large gap. Moreover, both ways will enhance the interlayer coupling and therefore lessen the lattice mismatch which is detrimental to gap opening.

\section{Conclusions~\label{conclusion}}

In conclusion, the experimentally observed metallic state in graphene/h-BN heterostructure is always attributed to the lattice
mismatch~\cite{Xue2011,Roth2013,Sachs2011,Titov2014}, and the electronic interaction is widely believed to enhance the gapped state~\cite{Hunt2013,Chen2014,Song2013,Bokdam2014,Jung2015}.
However, here we show that even without lattice mismatch, dominant onsite Coulomb repulsion of realistic value will suppress the band gap due to the
effective restoration of inversion symmetry between
neighboring carbons. The interaction-induced insulator-to-metal transition should also happen even if lattice mismatch is involved since
the phase diagrams are qualitatively the same despite different lattice configurations. Finally, we argue that the phenomenon of band gap suppressed by many-body effect should be present in other van der Waals heterostructures as long as a band gap induced by lattice asymmetry has already been opened in the noninteracting case.

\section{Acknowledgements~\label{Acknowledgements}}

We thank W. Ku for discussions. This work is supported by National Natural Science Foundation of China (No. 11474217), Program for New Century Excellent Talents in University (NCET-13-0428), and the Program for Professor of Special Appointment (Eastern Scholar) at Shanghai Institutions of Higher Learning as well as the Scientific Research Foundation for the Returned Overseas Chinese Scholars, State Education Ministry. J.-R. is also supported by Educational Commission of Anhui Province of China (No. KJ2013B059) and Natural Science Foundation of Anhui Province (No. 1308085QA05).

\appendix

\section{Solution of Coherent potential approximation to the model~(\ref{fullmodel})~\label{GF}}

The CPA average Green's function can be written in a matrix form
\begin{equation}
\begin{aligned}
\bar{G}^{-1}(\mathbf{k},\omega)\!&=\!\omega + \mu - \widehat{H_{tb}}(\mathbf{k}) -
&
\begin{bmatrix}
\begin{smallmatrix}
\Sigma_{C_1}(\omega) & 0               & 0          & 0   \\
0            & \Sigma_{C_2}(\omega)    & 0          & 0   \\
0            & 0               & 0          & 0   \\
0            & 0               & 0          & 0
\end{smallmatrix}
\end{bmatrix}
\end{aligned}
\end{equation}
where $\mu$ is chemical potential and $\widehat{H_{tb}}(\mathbf{k})$ is a $4\times4$ matrix obtained by applying the Fourier transformation to the model (\ref{fullmodel}). $\Sigma_{C_1}(\omega)$ and $\Sigma_{C_2}(\omega)$ represent for self-energies on two neighboring carbons. Now all spin indices have been omitted as we are interested in the paramagnetic phase. In real space, we have
\begin{equation}
\bar{G}_{ii}(\omega)=\frac{1}{\Omega_{BZ}}\int_{BZ}d\mathbf{k} \bar{G}_{ii}(\mathbf{k},\omega),
\label{realGreen}
\end{equation}
where the integral is over the first Brillouin zone of the sublattice and $i\in C,B,N$. Then a cavity Green's function $\mathcal
{G}_{i}(\omega)$ can be obtained through the Dyson equation
\begin{equation}
\mathcal {G}_{i}^{-1}(\omega)=\bar{G}_{ii}^{-1}(\omega)+\Sigma_{i}(\omega)
\end{equation}
for sublattice ($i\in C_1,C_2$) in graphene layer, which describes a medium with self-energy at a chosen site removed. The cavity can now be filled by a real "impurity" with disorder potential, resulting in an impurity Green's function
\begin{equation}
G_{i}^{\gamma}(\omega)=[\mathcal {G}_{i}^{-1}(\omega)-E_{i}^{\gamma}]^{-1}
\end{equation}
with impurity configurations of
$E_{i}^{\gamma}=\begin{cases}U/2&\gamma=+\\
-U/2&\gamma=-\end{cases}$ as defined by Eq.(\ref{rand.potential}). The CPA requires
\begin{equation}
\langle G_{i}^{\gamma}(\omega)\rangle =\bar{G}_{ii}(\omega),
\label{aveGreen}
\end{equation}
where the average is taken over the impurity configuration probabilities defined by Eq.(~\ref{rand.potential}).

Equation (\ref{realGreen}) and (\ref{aveGreen}) need to be solved self-consistently. By adjusting the chemical potential $\mu$, the condition of $\sum\limits_{i\in C,B,N}\langle n_{i}\rangle=4$ can be satisfied for half-filled case, where
\begin{equation}
\langle n_{i}\rangle=-\frac{1}{\pi}\int_{-\infty}^{\mu}Im\bar{G}_{ii}d\omega .
\end{equation}
It is also necessary to ensure that the resulting integrated DOS for each site should be consistent with the average occupation number probabilities used in Eq.(\ref{aveGreen}), so an extra loop of self-consistency should be added.


\begin{thebibliography}{99}
\bibitem{CommentTB} Besides the semi-local approximations to the exchanges and correlations, density functional theory calculations (DFT) have also included the Hartree term of the nonlocal Coulomb interaction. Therefore, we mentioned that the effect of nonlocal terms of the Coulomb interactions has been implicitly incorporated in our study at the DFT level. However, the exact exchange term, i.e., the Fock term, is not involved in our DFT calculations based on local density approximation. As we know, the exact exchange term will enhance the band gap. While this will eventually lead to a shift of phase boundary, the overall phase diagram remains qualitatively unchanged.
\bibitem{CommentCPA} Though the coherent potential approximation (CPA) can successfully capture the metal-insulator transition, here we would like to make a brief summary of its limitations which are discussed detailedly in Refs.~[\onlinecite{RowlandsJPCM1,RowlandsJPCM2}]; the first is the inability to describe magnetism, and the second is the incapability to determine the nature of the metallic state. However, these limitations will not affect the applicability of CPA to our current research, since we are only interested in the nontrivial issue of whether there is interaction-induced insulator-metal transition possibly happening in graphene on hexagonal boron nitride where no long-range magnetic order is observed. Please note, it is not our attempt to solve the minor details of whether the metal-insulator transition is accompanied by the antiferromagnetic phase transition or whether the  metallic state is a Fermi liquid, a non-Fermi liquid, or a marginal Fermi liquid.
\end{thebibliography}
\end{document}